# Incremental learning of LSTM framework for sensor fusion in attitude estimation

Parag Narkhede[1], Rahee Walambe[2], Shashi Poddar[3], Ketan Kotecha[2]

[1] Symbiosis Institute of Technology, Symbiosis International (Deemed University), Pune, Maharashtra, India
[2] Symbiosis Centre for Applied Artificial Intelligence, Symbiosis International (Deemed University), Pune, Maharashtra, India
[3] Central Scientific Instruments Organisation, Council of Scientific and Industrial Research, Chandigarh, Chandigarh, India

## Abstract

This paper presents a novel method for attitude estimation of an object in 3D space by incremental learning of the Long-Short Term Memory (LSTM) network. Gyroscope, accelerometer, and magnetometer are few widely used sensors in attitude estimation applications. Traditionally, multi-sensor fusion methods such as the Extended Kalman Filter and Complementary Filter are employed to fuse the measurements from these sensors. However, these methods exhibit limitations in accounting for the uncertainty, unpredictability, and dynamic nature of the motion in real-world situations. In this paper, the inertial sensors data are fed to the LSTM network which are then updated incrementally to incorporate the dynamic changes in motion occurring in the run time. The robustness and efficiency of the proposed framework is demonstrated on the dataset collected from a commercially available inertial measurement unit. The proposed framework offers a significant improvement in the results compared to the traditional method, even in the case of a highly dynamic environment. The LSTM framework-based attitude estimation approach can be deployed on a standard AI-supported processing module for real-time applications.

## Introduction

Sensor fusion is the process of combining information from multiple sensors to provide an improved version of the output compared to that from an individual sensor (White, 1991). It provides the advantage of improved accuracy, increased precision and robustness. One of the relevant and important sensor fusion applications is attitude estimation, which plays a vital role in detecting the position and orientation of the moving body in 3D space. The devices consisting of gyroscope, accelerometer, and magnetometer sensors are together responsible for predicting the state of the system. The attitude estimation has proven to be an essential base for robotic guidance and control applications. The object's orientation in the 3-dimensional space is termed as the attitude of the object (Titterton, Weston & Weston, 2004) and is represented as the rotation about the three orthogonal axes as Roll, Pitch, and Yaw.

In order to compute the necessary attitude angles, the attitude estimator should be accurate and handle uncertainties and disturbances such as sudden accelerations, strong vibrations, sudden external forces, magnetic disturbances, etc. Such external and irregular disturbances affect the performance of the sensor, and as such, a reliable and robust computational system is essential for accurate predictions. A system such as an Unmanned Aerial Vehicle (UAV) or Automated Guided Vehicle (AGV) needs to act instantly even when the Global Positioning System (GPS) fails to provide the necessary positioning and attitude signals. In all such situations, a computationally efficient and easy to deploy attitude estimation system is extremely crucial.

To this end, the tri-axial inertial sensors (gyroscope and accelerometer) along with the magnetic sensor are employed in the attitude estimating system. This combined sensors unit is termed as an Inertial Measurement Unit (IMU). IMU is an integral part of any modern-day Attitude Heading and Reference System (AHRS). Algebraic integration of angular velocity (obtained by the gyroscope) is used to compute the attitude of the moving object. However, due to continuous integration over time, the estimations start drifting from their true values. Inherently, the gyroscope possesses a time-varying bias characteristics. These limitations hamper the use of gyroscope as a standalone device for attitude estimation application. To overcome the drawbacks of the individual gyroscope, the measurements from the accelerometer and magnetometer are fused with gyroscope (Fourati & Belkhiat, 2016; Gebre-Egziabher, Hayward & Powell, 2004). The accelerometer is a sensor that measures the linear acceleration along its sensitive axis, and the magnetometer measures the magnetic field strength present in the sensor's vicinity. With the advancement in Micro Electrical Mechanical System (MEMS) technology, these sensors are now available in tiny sizes with equivalent or improved performance than mechanical sensors. Although MEMS sensors offer high performance, they suffer from several errors such as sensor misalignment, bias instability, linearity, etc. (Aydemir & Saranlı, 2012). The gyroscope measurements are robust to high-frequency noise whereas the accelerometer, and magnetometer, perform well in the low-frequency motions dynamics (Tseng et al., 2011). Such complementary behavior of these sensors makes them appropriate for the sensor fusion application (Poddar et al., 2017; Narkhede et al., 2019).

Several techniques and methods are proposed in the literature for sensor fusion. Kalman filter (KF) (Kalman, 1960) is one of the most widely adopted sensor fusion methodologies for engineering systems. Due to the nonlinear relationship between the multiple sensor measurements, the nonlinear version of KF (Julier & Uhlmann, 2004), i.e., Extended Kalman Filter (EKF) and Unscented Kalman Filter (UKF) are popular in attitude estimation application (Crassidis, Markley & Cheng, 2007). A simple structured complementary filter is also used for sensor fusion in low-cost attitude estimation systems (Mahony, Hamel & Pflimlin, 2008). However, these filters require appropriate tuning of the parameters ($\alpha$ or $K_P$ and $K_I$ for CF whereas matrices Q and R for EKF). KF is a complex process and requires appropriate mathematical modeling of the systems and noise characteristics (Higgins, 1975). Artificial intelligence and machine learning-based fusion methods are also proposed for multiple engineering and non-engineering applications in the last few years. Recurrent Neural Network (RNN) (Rumelhart, Hinton & Williams, 1986) and Long Short-Term Memory (LSTM) (Hochreiter & Schmidhuber, 1997) are widely adopted sequence models when the data is in sequential form. These sequence models are particularly effective than the vanilla neural networks since they can handle the underlying sequential relationship in the time series data. In engineering applications, significant performance improvement can be observed in consumer devices, industrial devices, healthcare technologies, robotic applications, etc with the use of these learning architectures. It is also observed that the performance of sequence models is enhanced by incorporating the EKF frameworks with them (Wang & Huang, 2011; Vural & Kozat, 2019; Bao et al., 2020). With the advancements in UAV technologies, deep learning methods are widely adopted in UAV operations like object detection and classification, maintaining UAV Internet of Things (IoT) Networks, traffic flow monitoring (Maimaitijiang et al., 2020; Zhu, Qi & Feng, 2020; Al-Sharman et al., 2019), etc.

In general, when the deep learning frameworks are implemented for a specific application; the networks are trained offline using the available training data. In most of the applications considered in this paper, the data is highly dynamic due to the uncertain and unpredictable dynamic movements of the system in the air. The traditional offline training based models fail to provide accurate predictions in such changing scenarios. There is a need for online training of the network to incorporate the newly available unseen data and update the deep learning model. This enables learning and understanding of the dynamic motion, which is an important aspect considered while designing any navigation system. This particular challenge can be mitigated by the use of an incremental learning approach. This paper proposes the incremental learning of the LSTM network to yield attitude angles. The main contributions of this paper are two fold; firstly, we have demonstrated the use of the LSTM network for attitude estimation and compared its performance with the traditional approach, specifically EKF. Secondly, we have developed and demonstrated the incremental learning technique for attitude estimation while training the LSTM network intermittently during the run. This ensures that the model is robust against sudden changes and handles the unseen data common in such dynamic problems convincingly. The results are compared with the EKF and traditional LSTM technique to prove the efficacy of the proposed approach.

The remaining part of the paper is organized as follows: 'Theoretical background' provides the details on theoretical background relevant to understanding of the paper and relevant literature while 'Proposed methodology' presents the details of LSTM and incremental LSTM (LSTM–inc) framework for attitude estimation. 'Results and discussion' analyses the proposed LSTM–inc framework in comparison to the other techniques, and 'Conclusion' concludes the paper.

## Theoretical Background

Attitude estimation is the process of estimating the orientation of an object in the three-dimensional space. Inertial sensors viz gyroscope, accelerometer, and magnetometer are widely used in attitude estimation systems. The measurements from these sensors are combined using the sensor fusion techniques for accurate estimation of the attitude angles.

### Review of literature

Kalman filter (KF) and its nonlinear versions EKF and UKF are some of the widely popular sensor fusion frameworks used in engineering applications. These frameworks follow the two-step process consisting of prediction and correction (Kalman, 1960). Even though the KF and its different versions are very popular, they involve complex mathematical operations and are not suitable for resource constrained small systems. Also, the performance of these KF frameworks are highly dependent on the noise parameter values fed to the system model (Euston et al., 2008).

In literature, KF is found to be a successful data fusion methodology in various application domains ranging from medical (Li, Mark & Clifford, 2007), agricultural (Huang et al., 2007), manufacturing (Jemielniak & Arrazola, 2008), positioning, and navigation (Li et al., 2011; Kumar et al., 2017), etc. KF was first reported in 1960 as a sophisticated replacement for the complex instrumentation system of attitude estimation (Kalman, 1960). Numerous versions of KF and EKF are available in the literature for the task of accurate attitude estimation. However, the KF requires the appropriate modeling of the noise parameters for its efficient operation. These parameters need to be tuned appropriately for reliable estimation. Considering this, various adaptive filter-tuning methodologies are present in the literature. Fuzzy logic-based tuning of the EKF is one of the most popular methodologies (Shi, Han & Liang, 2009; Sasiadek & Wang, 2003; Jwo et al., 2013; Yazdkhasti & Sasiadek, 2018). Multiple model adaptive estimation technique was employed with the EKF for attitude estimation application by Kottath et al. (2016). An Evolutionary optimization-based EKF tuning framework is presented by Poddar et al. (2016).

Another framework of sensor fusion, known as Complementary filter (CF), is also very popular. CF is a simple structure comprising of a low pass and a high pass filter (Higgins, 1975) working in complementary mode. Researchers have also come up with the nonlinear version of the CF, which uses the Proportional Integral (PI) Controller as a low pass filter (Mahony, Hamel & Pflimlin, 2008). Both the versions of CF neither require any consideration of noise characteristics nor the system model. Although complementary filter is competitive to Kalman filter, it also requires adaptive and appropriate tuning of the filter parameters. CF based attitude estimation system developed by Mahony, Hamel & Pflimlin (2008) is amongst some of the popular frameworks. Several research has occurred in the area of complementary filter which aims at adapting the gain parameters of the complementary filter have been proposed in the literature in the past (Kottath et al., 2017; Narkhede et al., 2019; H, 2019). A cascaded structure combining linear and non-linear version of complementary filter is presented by Narkhede et al. (2021).

Although different versions of KF and CF exist in the literature, they still have their own limitations of tuning and complex mathematical operations. Both these frameworks need adaptive adjustment of the filter parameters for the efficient performance of the filters, which requires additional complex subroutine for filter parameter adaptation.

With the advancements in Artificial Intelligence-based techniques and the availability of complex processing devices, research in the area of deep learning (DL) is increasing. DL frameworks are also applied in UAVs for detection and classification purposes. Although Deep Learning techniques are widely used in UAV applications, they are not much explored in attitude estimation applications. Al-Sharman et al. (2019) proposed a deep learning methodology for understanding the noise characteristics and tune the noise variance parameters in the Kalman filter structure. Huang et al. (2007) devised a method of using Convolutional Neural Networks (CNN) for estimating vehicle attitude based on angular velocity and acceleration. An unsupervised mechanism of deep auto-encoders for attitude estimation is proposed by Dai et al. (2018). It uses ANN-based encoder and decoder models to construct the Auto-encoder structure. Long-Term Short-Term Memory (LSTM) neural network was proposed for estimating attitude angles in Liu, Zhou & Li (2018). Hussain et al. (2019) proposed an indoor positioning system that uses LSTM architecture for activity recognition and prediction. LSTM is also proposed for localization and the motion estimation applications in Guo & Sung (2020) and Zhang et al. (2021)

All these approaches proposed offline training of the deep learning models where models are trained offline using the available datasets. However, in real-world applications, the dataset is not static, and due to the changing and dynamic nature of the incoming data due to unpredictable motions of the aerial vehicle in flight, offline training is bound to fail. Consider a case where sufficient data for various cases of a maneuver of an aerial vehicle is collected, and an LSTM model is trained to predict the system state. However, when the aerial vehicle is operated in real-time, it is subjected to numerous dynamical changes and motion variations, which may not be present in the training data. Based on these real-time changes, the model has to still predict the best possible estimate, which may prove challenging and inaccurate at times. Deep learning models being data-agnostic, may fail to understand the dynamically changed and unknown sensor inputs. Hence, these models tend to generate inaccurate predictions of the attitude angles at times. This creates a need for updating the network based on real-time sensor inputs. An incremental learning system keeps on predicting the system state while updating the existing knowledge (weights) of the LSTM framework during the run. In this work, an incremental learning approach is therefore proposed for estimating attitude for a moving vehicle.

**Attitude estimation from inertial sensors**

The attitude angle or the orientation of any moving vehicle is defined as the angle that the body frame of the vehicle makes with the earth's reference frame (Titterton, Weston & Weston, 2004; Farrell, 2008). The three different attitude angles of any vehicle about the x, y, and z-axes are known as Roll ($\phi$), Pitch ($\theta$), and Yaw($\psi$), respectively. Euler angle representation of attitude angles is followed in this paper as it has a direct relation with the rotation angles making it easy and convenient for understanding. The angular velocities measured by a gyroscope in the body reference frame about x, y, and z axes are generally denoted by p, q, and r, respectively. The equivalent reference frame quantities for the measured body frame angles can be obtained using the standard coordinate frame transformations taken in sequence. The rotation sequence followed in aerospace applications is the rotation about the z-axis followed by a rotation about the y-axis and then rotation about the x-axis (Pedley, 2013). The Greek letters phi ($\psi$), theta ($\theta$), and psi ($\psi$) are traditionally used to represent the Euler angles Roll, Pitch, and Yaw, respectively.

The time derivative of Euler angles can be computed using Eqs. (1)–(3).

$$\dot{\phi} = p + q \sin(\phi) \tan(\theta) + r \cos(\phi) \tan(\theta) \quad (1)$$

$$\dot{\theta} = q \cos(\phi) - r \sin(\phi) \quad (2)$$

$$\dot{\psi} = q \sin(\phi) \sec(\theta) + r \cos(\phi) \sec(\theta) \quad (3)$$

Further, Euler rates ($\dot{\phi}, \dot{\theta}, \dot{\psi}$) can be integrated to obtain the attitude estimates from the gyroscope. As the integration is carried out as successive addition, it leads to the accumulation of the unwanted component in the measurements. As time increases, the drift increases, misleading the gyroscope to be used as a standalone attitude estimating device. Therefore, the fusion of another sensor's measurements to the gyroscope measurements is essential for the accurate attitude estimation.

If $a_x$, $a_y$, and $a_z$ denote the linear accelerations measured along three orthogonal axes, respectively. If the disturbance due to the external parameters is ignored, then the roll and pitch angles using accelerometer can be computed using Eqs. (4) and (5), respectively.

$$\phi_a = tan^{-1}\left(\frac{a_y}{a_z}\right) \qquad (4)$$

$$\theta_a = tan^{-1}\left(\frac{-a_x}{a_y \sin\phi + a_z \cos\phi}\right) \qquad (5)$$

If $m_x$, $m_y$, and $m_z$ represents the magnetic strengths along x, y, and z-axes measured using the tri-axial magnetometer, then the heading/yaw angle can be computed using a magnetometer by the Eq. (6).

$$\psi_m = \tan^{-1}\left(\frac{m_z \sin\phi - m_y \cos\phi}{m_x \cos\theta + m_y \sin\theta \sin\phi + m_z \sin\theta \cos\phi}\right) \qquad (6)$$

The frequency characteristics of these sensors are complementary, and the inability of the gyroscope to be used as a stand along with the device for orientation estimation, a fusion of these sensors is essential for accurate attitude estimations.

**Sequence models in deep learning**

Sequence models are typically used for handling sequential data such as sensor data, text, sound data, and specific data with underlying sequential structure. A sequential data can be used to carry out various applications, including financial time series prediction, text processing, name entity recognition, sequence value prediction, etc. Sequence models are used for time series data in either the input or output of which the Recurrent Neural Network (RNN) is a good example. The traditional neural network model, also called Artificial Neural Network (ANN), does not exhibit the property of looping and handling time dependencies between the data (Bengio, Simard & Frasconi, 1994) and is therefore not a preferred solution for time-series data. RNN has a loop in them and allow information to persist. RNN is not altogether different from the traditional ANN, whereas it can be considered as the multiple ANN models connected in a series and one model passing the information to its successor.

In RNN, the temporal dynamics are preserved via an internal memory. However, for longer sequences, RNN proves inadequate and hence a variant called Long-Short Term Memory (LSTM) is developed by Hochreiter & Schmidhuber (1997). LSTMs are a special type of RNN having memory elements inside them and having capabilities of maintaining long-term dependencies between the data. Like RNN, LSTM is also a chain of ANN structures in which each of the ANN structure has four layers interacting in a specific manner. The detailed working of the RNN, LSTM, is beyond the scope of this paper.

**Incremental modelling**

Incremental learning is the method in computer science where the input data is continuously used to update the existing knowledge of the system. Incremental learning refers to the learning of the model in real-time for any dynamic changes in the system state. This method aims to adopt the new data without affecting the existing understanding and knowledge of the system. Figure 1 shows the incremental learning methodology. As shown in the figure, the input is applied to the learning algorithm, and the prediction model is updated with the newly obtained parameter values. When the new type of input appears, the model is updated using the new inputs as well as previous predictions of the model. This reduces the learning time of the algorithm and only updates the already learned model as and when the new data is made available.

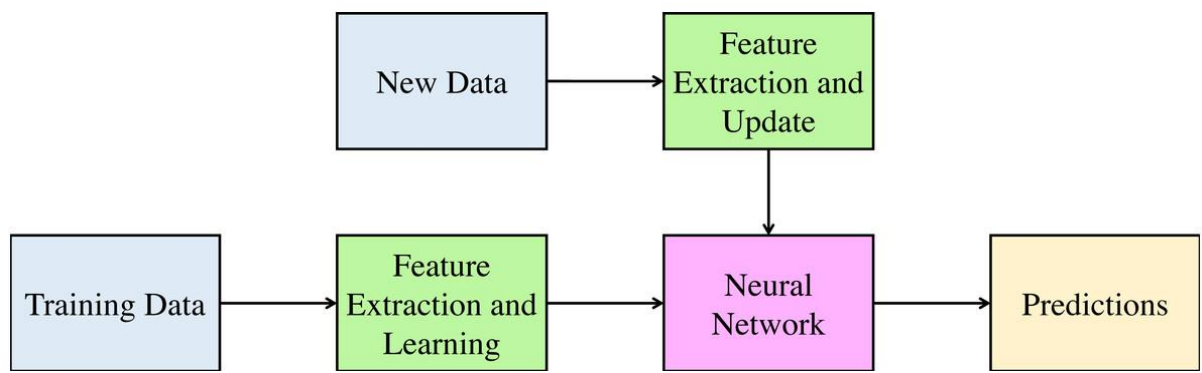

Figure 1: Incremental Learning Methodology

The following section provides detailed information regarding the proposed incremental learning methodology for the problem of attitude estimation.

**Proposed Methodology**

In the work presented in this paper, it is proposed to use the LSTM framework as a sensor fusion tool for the fusion of measurements from inertial sensors for accurate attitude angle estimation.

In this application of AHRS under consideration, data from the gyroscope, accelerometer, and magnetometer are considered for fusion. All these sensory data have an underlying sequential/ temporal dependency. Traditional ANN does not have a mechanism to handle these sequential dependencies, and hence they are insufficient for such tasks (Bengio, Simard & Frasconi, 1994). In RNN, the temporal dynamics are preserved via an internal memory. However, for longer sequences, RNN proves inadequate, and hence a variant called Long-Short Term Memory is usually used (Hochreiter & Schmidhuber, 1997). In this work, an LSTM based network is proposed, which can be applied to inertial sensor data for estimating the roll, pitch, and yaw angles. It is also proposed here to train the model during run-time in batches to become robust for the new real-time data. Figure 2 shows the proposed incrementally trained LSTM (hereafter referred as LSTM-inc) structure of sensor fusion for attitude estimation. The output of all three tri-axial sensors are applied as an input to the LSTM framework. All these nine measurements are concatenated together to form an array of input. Measurements from current step and previous step are used as input for estimating attitude angles of current step. This indicates the use of time-step as 2 while giving input to the LSTM layer. The implemented model consists of two hidden layers of LSTM units. The prediction is made using the neuron's layer with a linear activation function. The supervised mechanism of LSTM is used to learn the features from inputs and produce the desired output. The learning model is trained offline using a part of the available dataset. In the prediction phase, attitude angles are estimated based on the trained model and sensor measurements. After a specific time-interval, the weights of the trained model are again updated to learn the new features from the inputs, and attitude angles are predicted further. In this work, the interval of 30 s (3000 samples) is considered for weights updation and is carried out for the whole sequence. The step by step procedure for the proposed methodology is presented in the flowchart shown in Figs. 3 and 4. Phase 1, where the offline training of the network is carried out, is shown in Fig. 3; whereas the attitude estimation and incremental learning phase is presented in Fig. 4. The proposed attitude estimating framework uses measurements from the tri-axial accelerometer, tri-axial gyroscope, tri-axial magnetometer as its inputs and yields Roll, Pitch, Yaw angles as output.

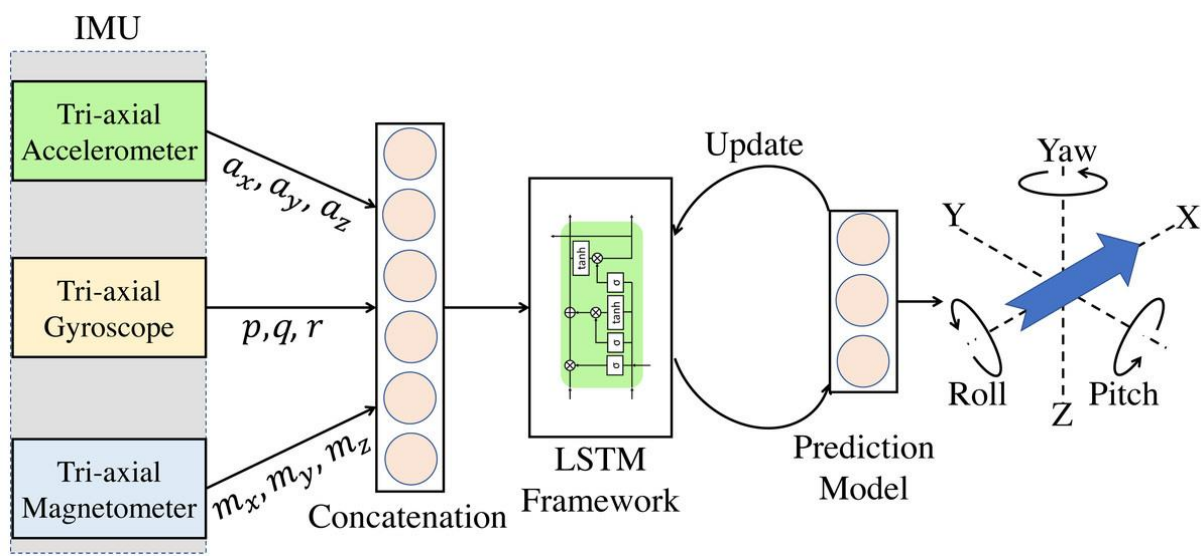

Figure 2 LSTM-inc: Proposed LSTM based incremental learning framework for attitude estimation.

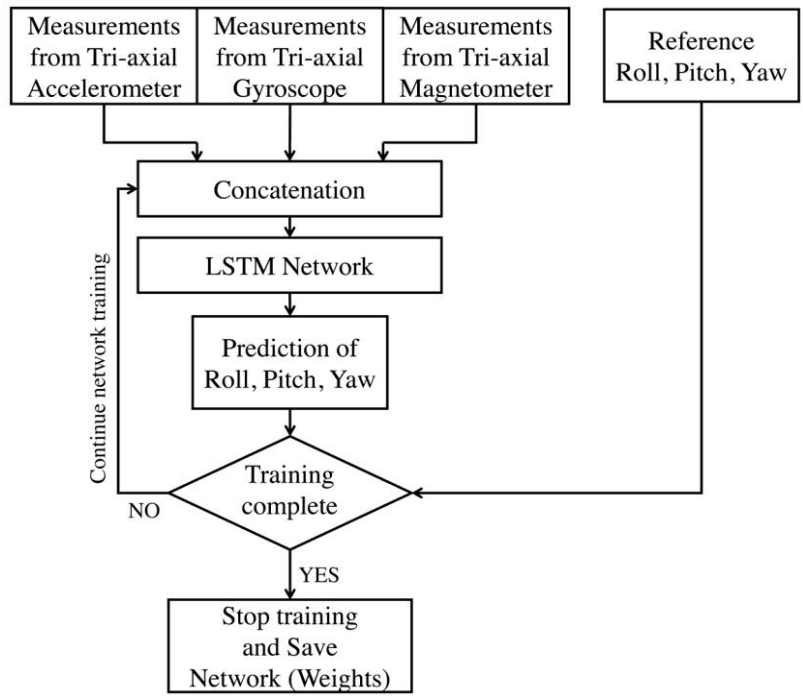

Figure 3 LSTM based sensor fusion for attitude estimation (offline training phase).

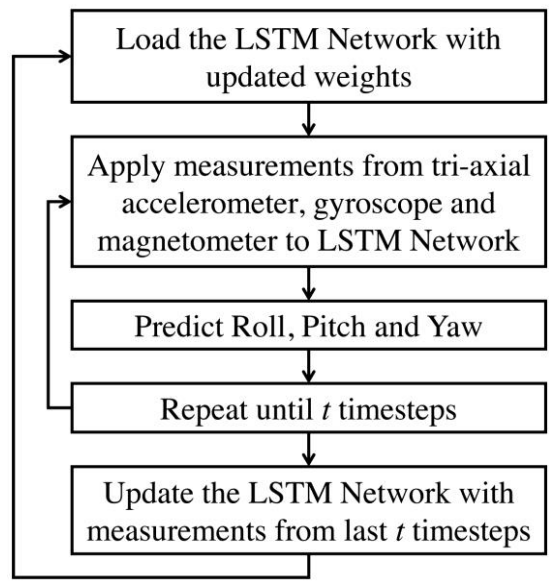

Figure 4 LSTM based sensor fusion for attitude estimation (estimation and incremental learning phase).

In this proposed framework, the sensors' measurements are applied as an input to the LSTM structure. However, in the traditional attitude estimation structures, the attitude angles are computed using individual sensors initially (using Eqs. (1)–(6)), and then combined together to yield estimated output. The direct application of sensor measurements as an input to the proposed framework avoids angle computation requirement using individual sensors thereby avoiding complex mathematical operations on raw sensor measurements. The online/ incremental learning of the LSTM neural network assures incorporation of newly arrived dynamic changes in the inputs and makes the estimation system robust.

The next section provides a detailed analysis and experimental proof for the proposed LSTM based sensor fusion framework of attitude estimation. The section also compares the proposed architecture with the traditional attitude estimation techniques.

## Results and Discussion

The proposed deep learning-based LSTM neural network structure for attitude estimation is applied to the different datasets collected from the commercially available AHRS module. EKF-based Xsens MTI-G module is considerably accurate and commercially available inertial measurement system (Xsens, 2020). It provides information about the raw sensor measurements from the tri-axial accelerometer, tri-axial gyroscope, and tri-axial magnetometer, as well as the reference orientation angles for the motion provided. The data is logged at the rate of 100 samples per second from the Xsens module. Random motions are given to this module, and the data is recorded for offline simulations and analysis. As the datasets are logged using the actual IMU, the generated data is considered as real-world data. These logged attitude angles from the Xsens module are considered as ground truth in the supervised learning framework. The raw sensor measurements are applied to the proposed LSTM-inc framework to yield attitude angles. Root Mean Square Error (RMSE) is considered for quantitatively verifying the feasibility and accuracy of the proposed structure with respect to the considered reference. For N samples under considerations, if x is the ground truth and $\hat{x}$ being the predictions, then the RMSE can be calculated using the Eq. (7).

$$RMSE = \sqrt{\frac{\sum_{i=1}^{N}(x_i - \hat{x}_i)^2}{N}} \quad (7)$$

In the first phase of experimentation, the LSTM model is trained offline using the available data and tested on the testing dataset. Since the vehicular motion data is dynamic and indeterministic, it is required to incorporate the run time changes to train & update the LSTM model in an incremental fashion. Table 1 shows the comparison between the offline trained LSTM model (where incremental training is not involved) with the incrementally trained LSTM model. The comparison with the EKF is also presented for reference.

Table 1 RMSE comparison for the LSTM trained incrementally (LSTM-inc), LSTM trained offline (LSTM), and EKF (all values in radians).

| Dataset | Roll RMSE | | | Pitch RMSE | | | Yaw RMSE | | | Average RMSE | | |
| --- | --- | --- | --- | --- | --- | --- | --- | --- | --- | --- | --- | --- |
| | LSTM_inc | LSTM | EKF | LSTM_inc | LSTM | EKF | LSTM_inc | LSTM | EKF | LSTM_inc | LSTM | EKF |
| D1 | 0.05 | 0.09 | 0.02 | 0.03 | 0.05 | 0.02 | 0.12 | 0.22 | 0.18 | 0.07 | 0.12 | 0.07 |
| D2 | 0.04 | 0.05 | 0.02 | 0.03 | 0.05 | 0.02 | 0.07 | 0.07 | 0.08 | 0.04 | 0.06 | 0.04 |
| D3 | 0.05 | 0.06 | 0.02 | 0.03 | 0.03 | 0.03 | 0.11 | 0.16 | 0.13 | 0.06 | 0.09 | 0.06 |
| D4 | 0.33 | 0.35 | 0.74 | 0.06 | 0.04 | 0.04 | 0.31 | 0.35 | 0.94 | 0.23 | 0.24 | 0.58 |
| D5 | 1.26 | 1.33 | 1.14 | 0.05 | 0.08 | 0.05 | 0.35 | 0.34 | 0.87 | 0.55 | 0.58 | 0.68 |
| D6 | 1.52 | 1.76 | 2.46 | 0.45 | 1.10 | 0.49 | 1.17 | 1.40 | 2.53 | 1.05 | 1.42 | 1.83 |

In the Table 1, D1–D6 are the datasets collected from the Xsens module. The datasets D1-D4 are datasets with less dynamic motion, and D5-D6 are datasets involving higher dynamic motions. Roll RMSE columns indicate the RMSEs obtained for incrementally trained LSTM, offline LSTM and EKF; similarly, RMSEs for Pitch and Yaw are indicated. Average RMSE is the mean of RMSEs in Roll, Pitch, and Yaw angles. Improvement in the RMSEs is observable in the incrementally trained LSTM model. In case of less dynamic motions, even though the incrementally trained LSTM is not outperforming EKF estimates, the difference is not significant. However, substantial improvements is observable in the majority of the high dynamic motion cases. Table 2 indicates the variance of the Roll, Pitch and Yaw data for various data sets.

**Table 2: Variance of data.**

| Dataset | Roll | Pitch | Yaw |
| --- | --- | --- | --- |
| D1 | 0.173 | 0.079 | 0.049 |
| D2 | 0.144 | 0.064 | 0.054 |
| D3 | 0.140 | 0.081 | 0.042 |
| D4 | 0.051 | 0.002 | 0.052 |

| Dataset | Roll | Pitch | Yaw |
|---|---|---|---|
| D5 | 1.341 | 0.003 | 0.139 |
| D6 | 4.215 | 0.954 | 1.834 |

In the second phase of experimentation, the proposed incrementally trained LSTM framework is compared with the traditionally available and widely used Extended Kalman Filter and is shown in Figs. 5 and 6. Figure 5 shows the RMSE comparison of the proposed incrementally trained LSTM model with the EKF having the datasets involving fewer dynamic motions. Figure 6 indicates a similar comparison; however, the datasets involve higher dynamic motions characteristics.

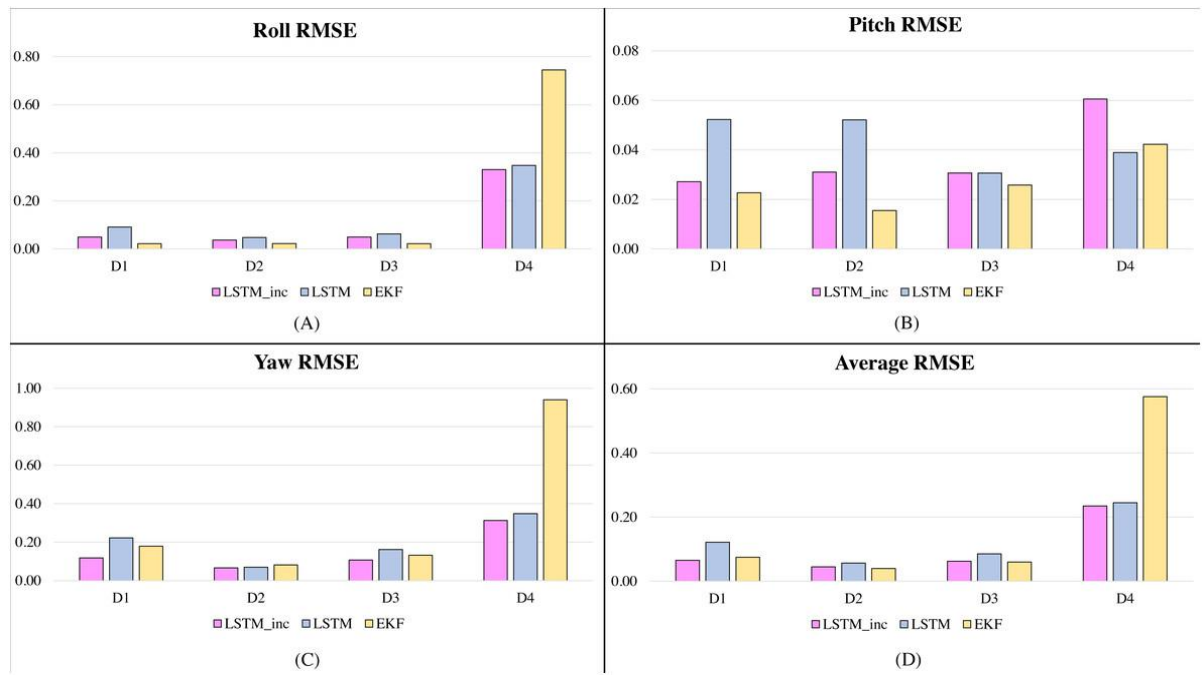

Figure 5 RMSE comparison of proposed incrementally trained LSTM model with EKF (less dynamic motion datasets): (A) RMSE in Roll, (B) RMSE in Pitch, (C) RMSE in Yaw, (D) Average RMSE.

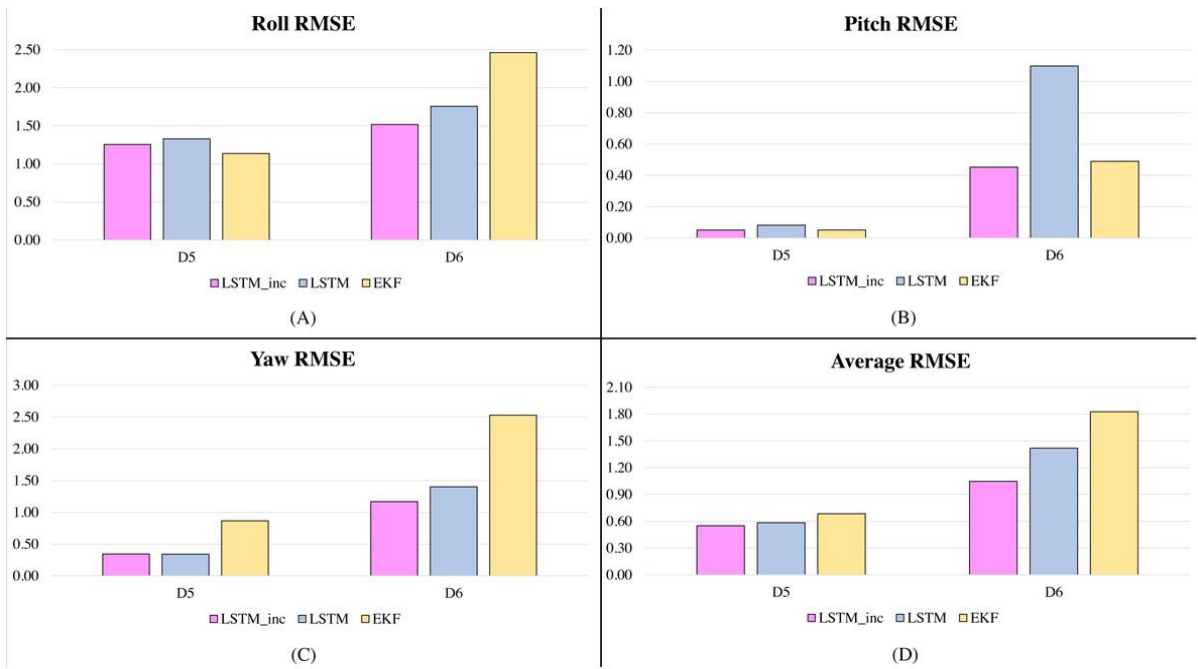

Figure 6 RMSE comparison of proposed incrementally trained LSTM model with EKF (high dynamic motion datasets): (A) RMSE in Roll, (B) RMSE in Pitch, (C) RMSE in Yaw, (D) Average RMSE.

The sample estimation results are shown in Figs. 7 and 8. These figures indicate the reference attitude angles (Legend: Reference) as well as the estimated attitude angles using an incrementally trained LSTM framework (Legend: LSTM-inc). The estimates of offline trained LSTM framework (Legend: LSTM) are also represented for comparison. The estimated output through EKF is also indicated in figures (Legend: EKF). Figure 7 shows the attitude estimates for the less dynamic motion dataset, whereas high dynamic changes in the attitude estimates are observable in Fig. 8.

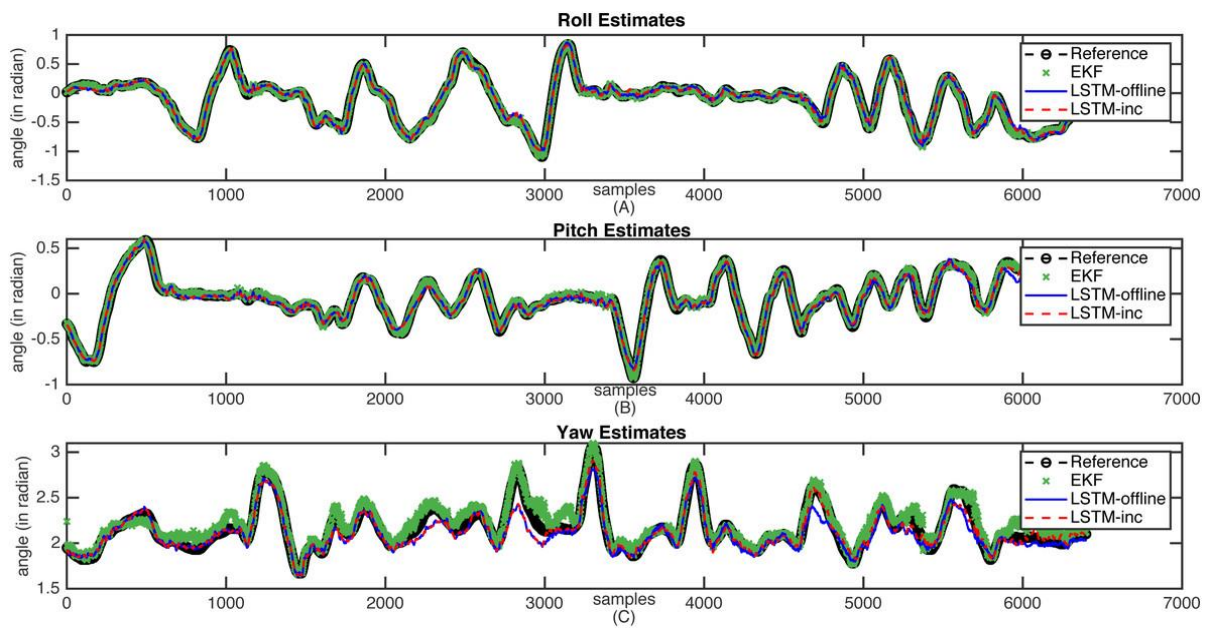

Figure 7 Sample attitude angle estimates (less dynamic motion): (A) Roll estimates, (B) Pitch estimates, (C) Yaw estimates.

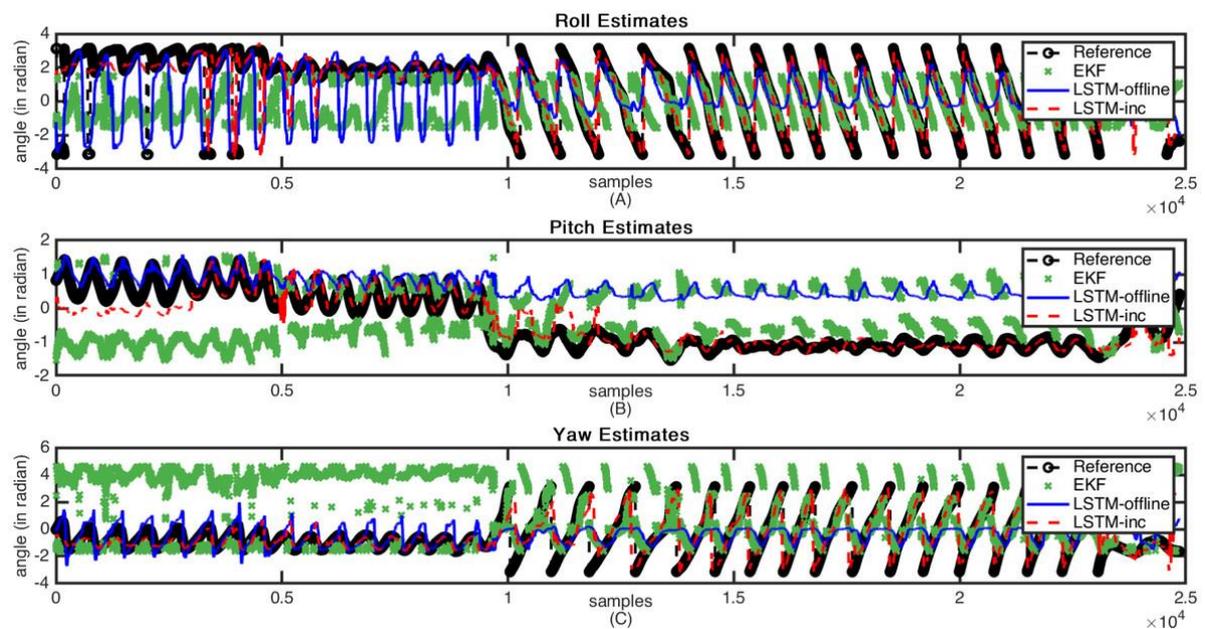

Figure 8 Sample attitude angle estimates (high dynamic motion): (A) Roll estimates, (B) Pitch estimates, (C) Yaw estimates.

From these figures, it can be observed that when there is a slow motion that does not have dynamic changes, the estimates of offline trained LSTM are almost equal to the incrementally trained LSTM. Whereas, in the case where motion involves dynamic changes, the offline trained LSTM fails to predict accurate estimations due to the lack of prior knowledge of the framework. However, in such cases, incrementally trained LSTM accommodates the changes that happened recently and provides accurate attitude angles. In the incremental training of the model, the LSTM model gets updated with the new type of inputs. The updation of the model weights helps in predicting the accurate attitude angles. The comparative analysis carried out with the traditional EKF based frameworks also shows that the proposed incrementally trained LSTM framework is able to predict a relatively accurate estimation as compared to the offline LSTM. The results for the presented EKF can be improved by incorporating adaptive tuning mechanism with it. However, this will again add to the computational complexity of the EKF. For comparison, the EKF parameters are tuned manually using trial and error method, and the analysis is carried out. The proposed methodology updates the LSTM model at regular interval, adding robustness to the estimation architecture. These analysis have helped in proving the feasibility of using the LSTM model for performing sensor fusion in attitude estimation applications, even in applications where motions are unpredictable and involve dynamic changes.

In the current work, the time interval of 30 s is considered for model weight update. However, in specific applications where a fixed type of motion is involved, the time interval can be increased to reduce the computational load. Due to the availability of dedicated AI processing hardware platforms, the weight update time interval can be reduced to incorporate every small change in the motions instantly. Accurate attitude estimation is security essential aspect. Due to automated feature extraction and learning in the LSTM framework, it may be difficult to understand the situations where the considered model will fail to predict accurate estimates. The issue can be resolved by considering the explainability of the models as explained in Le et al. (2019) and Dang et al. (2020). This will help to predict accurate angles and accordingly suggest the corrective actions making model more robust and will be studied in future.

## Conclusion

The LSTM based incremental learning framework for performing sensor fusion is presented in this work. The data from inertial sensors is fed to the LSTM architecture for reliable attitude estimation in the navigational framework. While the navigation system undergoes dynamic motions, the traditional attitude estimation techniques is not very accurate in following the trajectory and hence an incremental learning framework for the LSTM network is developed and demonstrated here in this paper. The proposed approach is evaluated with the help of real-world data sets collected from commercially available AHRS modules. These results are also compared with the traditional EKF framework and it is observed that the LSTM-inc method outperforms the EKF method specifically in the situations where dynamic motion changes are involved. Additionally, these results are compared with the offline trained LSTM and the comparison results demonstrate that the offline trained LSTM network fails to provide accurate estimation when a system undergoes an unknown type of motion. The nature of motion is generally dynamic, and it is not practically possible to obtain datasets involving all types of motions to train the LSTM network model in an offline mode. Hence, an incrementally trained framework is suitable for such applications. In future work, the relation between the multiple tri-axial sensors can be analyzed so that the redundant features (measurements) can be discarded while training the network. Along with the optimal feature selection, features can be weighted, and the results can be analyzed for improvement in estimation accuracy.

## Data Availability

The following information was supplied regarding data availability:

The code and raw data are available at GitHub: https://github.com/nsparag/LSTM-INC.git.

## References

**Al-Sharman MK, Zweiri Y, Jaradat MAK, Al-Husari R, Gan D, Seneviratne LD. 2019**. Deep-learning-based neural network training for state estimation enhancement: application to attitude estimation. IEEE Transactions on Instrumentation and Measurement **69**(1):24-34

**Aydemir GA, Saranlı A. 2012**. Characterization and calibration of mems inertial sensors for state and parameter estimation applications. Measurement **45**(5):1210-1225

**Bao A, Gildin E, Huang J, Coutinho EJR. 2020**. Data-driven end-to-end production prediction of oil reservoirs by enkf-enhanced recurrent neural networks. In: SPE latin american and caribbean petroleum engineering conference. Society of Petroleum Engineers.

**Bengio Y, Simard P, Frasconi P. 1994**. Learning long-term dependencies with gradient descent is difficult. IEEE Transactions on Neural Networks **5**(2):157-166

**Crassidis JL, Markley FL, Cheng Y. 2007**. Survey of nonlinear attitude estimation methods. Journal of Guidance, Control, and Dynamics **30**(1):12-28

**Dai X, Zhou Y, Meng S, Wu Q. 2018**. Unsupervised feature fusion combined with neural network applied to uav attitude estimation. In: 2018 IEEE international conference on robotics and biomimetics (ROBIO). Piscataway. IEEE. 874-879

**Dang T, Van H, Nguyen H, Pham V, Hewett R. 2020**. Deepvix: explaining long short-term memory network with high dimensional time series data. In: Proceedings of the 11th international conference on advances in information technology, IAIT2020. New York, NY, USA. Association for Computing Machinery.

**Euston M, Coote P, Mahony R, Kim J, Hamel T. 2008**. A complementary filter for attitude estimation of a fixed-wing uav. In: 2008 IEEE/RSJ international conference on intelligent robots and systems. Piscataway. IEEE. 340-345

**Farrell J. 2008**. Aided navigation: GPS with high rate sensors. United States: McGraw-Hill, Inc.

**Fourati H, Belkhiat DEC. 2016**. Multisensor attitude estimation: fundamental concepts and applications. Boca Raton: CRC Press.

**Gebre-Egziabher D, Hayward RC, Powell JD. 2004**. Design of multi-sensor attitude determination systems. IEEE Transactions on Aerospace and Electronic Systems **40**(2):627-649

**Guo H, Sung Y. 2020**. Movement estimation using soft sensors based on bi-lstm and two-layer lstm for human motion capture. Sensors **20**(6):1801

**Kazdal H. 2019**. Fuzzy based complementary filter design for attitude estimation. PhD thesis, Ankara Yıldırım Beyazıt Üniversitesi Fen Bilimleri Enstitüsü thesis

**Higgins WT. 1975**. A comparison of complementary and kalman filtering. IEEE Transactions on Aerospace and Electronic Systems **3**:321-325

**Hochreiter S, Schmidhuber J. 1997**. Long short-term memory. Neural Computation **9**(8):1735-1780

**Huang Y-B, Lan Y-B, Hoffmann W, Lacey R. 2007**. Multisensor data fusion for high quality data analysis and processing in measurement and instrumentation. Journal of Bionic Engineering **4**(1):53-62

**Hussain G, Jabbar MS, Cho J-D, Bae S. 2019**. Indoor positioning system: a new approach based on lstm and two stage activity classification. Electronics **8**(4):375

**Jemielniak K, Arrazola P. 2008**. Application of ae and cutting force signals in tool condition monitoring in micro-milling. CIRP Journal of Manufacturing Science and Technology **1**(2):97-102

**Julier SJ, Uhlmann JK. 2004**. Unscented filtering and nonlinear estimation. Proceedings of the IEEE **92**(3):401-422

**Jwo D-J, Yang C-F, Chuang C-H, Lee T-Y. 2013**. Performance enhancement for ultra-tight gps/ins integration using a fuzzy adaptive strong tracking unscented kalman filter. Nonlinear Dynamics **73**(1-2):377-395

**Kalman RE. 1960**. A new approach to linear filtering and prediction problems. Transaction of the ASME-Journal of Basic Engineering 35-45

**Kottath R, Narkhede P, Kumar V, Karar V, Poddar S. 2017**. Multiple model adaptive complementary filter for attitude estimation. Aerospace Science and Technology **69**:574-581

**Kottath R, Poddar S, Das A, Kumar V. 2016**. Window based multiple model adaptive estimation for navigational framework. Aerospace Science and Technology **50**:88-95

**Kumar GA, Patil AK, Patil R, Park SS, Chai YH. 2017**. A lidar and imu integrated indoor navigation system for uavs and its application in real-time pipeline classification. Sensors **17**(6):1268

**Le DD, Pham V, Nguyen HN, Dang T. 2019**. Visualization and explainable machine learning for efficient manufacturing and system operations. Smart and Sustainable Manufacturing Systems **3**(2):127-147

**Li Q, Mark RG, Clifford GD. 2007**. Robust heart rate estimation from multiple asynchronous noisy sources using signal quality indices and a kalman filter. Physiological Measurement **29**(1):15

**Li T-HS, Su Y-T, Liu S-H, Hu J-J, Chen C-C. 2011**. Dynamic balance control for biped robot walking using sensor fusion, kalman filter, and fuzzy logic. IEEE Transactions on Industrial Electronics **59**(11):4394-4408

**Liu Y, Zhou Y, Li X. 2018**. Attitude estimation of unmanned aerial vehicle based on lstm neural network. In: 2018 international joint conference on neural networks (IJCNN). Piscataway. IEEE. 1-6

**Mahony R, Hamel T, Pflimlin J-M. 2008**. Nonlinear complementary filters on the special orthogonal group. IEEE Transactions on Automatic Control **53**(5):1203-1218

**Maimaitijiang M, Sagan V, Sidike P, Hartling S, Esposito F, Fritschi FB. 2020**. Soybean yield prediction from uav using multimodal data fusion and deep learning. Remote Sensing of Environment **237**:111599

**Narkhede P, Joseph Raj AN, Kumar V, Karar V, Poddar S. 2019**. Least square estimation-based adaptive complimentary filter for attitude estimation. Transactions of the Institute of Measurement and Control **41**(1):235-245

**Narkhede P, Poddar S, Walambe R, Ghinea G, Kotecha K. 2021**. Cascaded complementary filter architecture for sensor fusion in attitude estimation. Sensors **21**(6):1937

**Pedley M. 2013**. Tilt sensing using a three-axis accelerometer. Freescale Semiconductor Application Note **1**:2012-2013

**Poddar S, Hussain S, Ailneni S, Kumar V, Kumar A. 2016**. Tuning of gps aided attitude estimation using evolutionary algorithms. International Journal of Intelligent Unmanned Systems **4**(1):23-42

**Poddar S, Narkhede P, Kumar V, Kumar A. 2017**. Pso aided adaptive complementary filter for attitude estimation. Journal of Intelligent & Robotic Systems **87**(3–4):531-543

**Rumelhart DE, Hinton GE, Williams RJ. 1986**. Learning representations by back-propagating errors. Nature **323**(6088):533-536

**Sasiadek J, Wang Q. 2003**. Low cost automation using ins/gps data fusion for accurate positioning. Robotica **21**(3):255-260

**Shi Y, Han C, Liang Y. 2009**. Adaptive ukf for target tracking with unknown process noise statistics. In: 2009 12th international conference on information fusion. 1815-1820

**Titterton D, Weston JL, Weston J. 2004**. Strapdown inertial navigation technology. Herts: IET. **Vol. 17**

**Tseng SP, Li W-L, Sheng C-Y, Hsu J-W, Chen C-S. 2011**. Motion and attitude estimation using inertial measurements with complementary filter. In: 2011 8th asian control conference (ASCC). Piscataway. IEEE. 863-868

**Vural NM, Kozat SS. 2019**. An efficient ekf based algorithm for lstm-based online learning. CoRR.

**Wang X, Huang Y. 2011**. Convergence study in extended kalman filter-based training of recurrent neural networks. IEEE Transactions on Neural Networks **22**(4):588-600

**White FE. 1991**. Data fusion lexicon. Technical report. Joint Directors of Labs Washington DC

**Xsens. 2020**. MTi User Manual. Xsens Technologies. (accessed February 2020) **software**

**Yazdkhasti S, Sasiadek JZ. 2018**. Multi sensor fusion based on adaptive kalman filtering. In: Advances in aerospace guidance, navigation and control. Cham: Springer. 317-333

**Zhang M, Jia J, Chen J, Deng Y, Wang X, Aghvami AH. 2021**. Indoor localization fusing wifi with smartphone inertial sensors using lstm networks. IEEE Internet of Things Journal Epub ahead of print 2021 19 March

**Zhu X, Qi F, Feng Y. 2020**. Deep-learning-based multiple beamforming for 5g uav iot networks. IEEE Network **34**(5):32-38